\pdfoutput=1 
\documentclass[12pt, a4paper]{article}

\usepackage[utf8]{inputenc}
\usepackage[T1]{fontenc}
\usepackage[margin=2cm]{geometry} 
\usepackage{lmodern}
\usepackage{parskip} 
\usepackage{amsmath,amssymb}
\usepackage{textcomp}

\usepackage{graphicx}
\usepackage{float} 
\usepackage[font=footnotesize,labelsep=period]{caption} 
\usepackage{booktabs}
\usepackage{array}

\makeatletter
\def\maxwidth{\ifdim\Gin@nat@width>\linewidth\linewidth\else\Gin@nat@width\fi}
\def\maxheight{\ifdim\Gin@nat@height>\textheight\textheight\else\Gin@nat@height\fi}
\makeatother
\setkeys{Gin}{width=\maxwidth,height=\maxheight,keepaspectratio}

\usepackage{xurl}
\usepackage{hyperref}
\hypersetup{
  hidelinks,
  pdfcreator={LaTeX}
}

\title{\textbf{\Large Autonomous Cyber Defense: Real-Time Attack Detection and Mitigation in Software-Defined Networks Using Machine Learning}}

\author{
  \begin{minipage}[t]{0.32\textwidth}
    \centering\small
    \textbf{Alexandre Aguiar Amaral} \\
    Department of Informatics \\
    Instituto Federal Catarinense \\
    Camboriú, Brazil \\
    alexandre.amaral@ifc.edu.br
  \end{minipage}\hfill
  \begin{minipage}[t]{0.32\textwidth}
    \centering\small
    \textbf{Fernando Luiz Moro} \\
    Department of Informatics \\
    Instituto Federal Catarinense \\
    Camboriú, Brazil \\
    fnnmoro@gmail.com
  \end{minipage}\hfill
  \begin{minipage}[t]{0.32\textwidth}
    \centering\small
    \textbf{Ana Paula Malheiro} \\
    Department of Mathematics \\
    Instituto Federal Catarinense \\
    Camboriú, Brazil \\
    ana.amaral@ifc.edu.br
  \end{minipage}
}
\date{}

\begin{document}

\maketitle

\begin{abstract}
\noindent Autonomous response has evolved into a timing-critical challenge rather than solely a matter of detection accuracy. In recent intrusions, the interval between initial access and the first lateral movement has been observed to be as short as 27 seconds, a window that precludes any human-in-the-loop workflow. This paper presents a closed-loop framework that detects and blocks attacks in software-defined networks without operator involvement, evaluating its performance against this stringent temporal constraint rather than relying exclusively on detection accuracy. An automated data pipeline collects IP flows and aggregates them into labeled training data, while a prevention module selects and trains candidate classifiers and issues blocking rules directly to the SDN controller. In a SYN flooding denial of service case study, the deployed K-Nearest Neighbors classifier achieved an F1 score of 96.7\% and the cycle from flow availability to enforced block completed in 21 seconds, below the fastest breakout time reported to date.

\par\vspace{\baselineskip}
\end{abstract}

\section{Introduction}\label{sec:introduction}

The evolution of information and communication technology has undeniably brought many benefits and conveniences. People and objects are increasingly connected to the Internet, while countless services once unimaginable have been developed. However, the complexity of managing computer networks has become evident, which also exposes the weaknesses inherent to the methods and solutions traditionally used for this purpose \cite{wu2025}.

New requirements and higher legal standards related to data security, privacy, quality, and availability of the services provided have been imposed. Quality and performance measures such as SLA (\emph{Service Level Agreement}), QoS (\emph{Quality of Service}), and QoE (\emph{Quality of Experience}) are increasingly demanded and, when not met, may lead to serious financial and reputational losses for the institution \cite{safdar2018}.

The genesis of IP networks did not account for the connection of billions of machines and users or for a myriad of services. Likewise, security mechanisms were not natively specified and, even after decades, there is still no definitive solution to address them. At the same time, networks have become more heterogeneous over the years (\emph{e.g.}, equipment, protocols), and the growing number of hardware and software vulnerabilities, together with the value of the information in transit, has increased the growth rate of attacks against security and privacy \cite{wu2025}, \cite{ayoubi2018}.

The rapid growth of the IoT (\emph{Internet of Things}) has also contributed to the use of a wide range of devices as weapons for attacks such as DDoS (\emph{Distributed Denial of Service}) \cite{wu2025}, \cite{preimesberger2018}, \cite{vaarandi2013}, \cite{amaral2013}. The heterogeneity and vulnerability of these network devices are obstacles to attack detection and mitigation, a problem that remains current and continues to motivate proposals targeted at home and edge environments \cite{mendonca2025}, \cite{raja2026}.

This compression of the attack timeline is measurable: breakout time, defined as the interval between initial access and the moment the adversary moves laterally to a second host, fell to an average of 29 minutes in 2025, a 65\% increase in speed over the previous year, with the fastest observed case at 27 seconds and data exfiltration starting four minutes after initial access in one intrusion \cite{crowdstrike2026}. Any response cycle that depends on an operator reading logs and manually pushing a rule is, by construction, slower than the adversary, and every minute spent on manual triage is a minute in which the intrusion may spread beyond the compromised host.

Network management tasks, in particular attack diagnosis and mitigation, are notoriously nontrivial. Mechanisms that cannot adapt to the evolution and nuances of networks, and that require fully manual operation, become increasingly ineffective. During an attack, \emph{logs} with gigabytes of data (\emph{e.g.}, alarms) may be generated \cite{safdar2018}. Decision making requires the analysis of the generated data, which is a slow and costly process when performed manually \cite{vaarandi2013}, \cite{amaral2012}. Depending on the network and the application, a few minutes of downtime may cause substantial losses.

These observations have led to the search for new network management strategies. In this scenario, mechanisms are needed that enable more effective management, reducing manual intervention and actions, and that are able to adapt to the evolution of networks and of the strategies used in attacks \cite{amaral2017}. In this context, software-defined networking (SDN) and machine learning algorithms emerge as alternatives to address the challenges described above \cite{wu2025}, \cite{ayoubi2018}.

The emergence of SDN has been crucial, since it separates the control and forwarding functions in network devices, allowing management to be directly programmable \cite{onf}. An SDN can be implemented through OpenFlow technology, which is an open standard \cite{ayoubi2018}. It provides a standardized programming interface that allows direct control of network forwarding devices. The main advantage of SDN is that administration and management tasks are simplified and performed in a homogeneous way, which mitigates the complexity found in conventional networks \cite{ahmad2015}.

Artificial intelligence plays a crucial role in monitoring and securing the network devices, leveraging machine learning to detect attacks and respond to threats in real-time \cite{ahmad2015}, \cite{cisco2018}. Machine learning allows a computer to learn without being explicitly programmed \cite{das2017}. Its methods make it possible to automatically extract patterns from network data, learn from them, and detect known attacks, as well as to support the diagnosis of new attacks \cite{najafabadi2015}.

Many machine learning algorithms have been proposed in the literature \cite{ayoubi2018}. Nevertheless, several challenges still limit their wider adoption. Among them are the aspects related to modeling the problem to be addressed, to how the algorithms are trained and evaluated, and to how these processes can be automated.

This work proposes a machine learning based system that monitors network traffic in real time, diagnosing attacks and automatically applying countermeasures when an attack occurs. The modeling, training, and evaluation of different algorithms are automated by the system. It also provides the convenience and flexibility of choosing which algorithm is most suitable for the monitored network environment, based on several effectiveness and efficiency metrics. In addition, it allows the network administrator to specify which automated actions are to be executed in a software-defined network when an attack occurs.

The proposed system seeks to speed up the tasks of detecting attacks and triggering corrective actions, minimizing manual intervention. We also believe in its potential to provide greater network resilience and to maximize the availability and reliability of the services it delivers. In this way, the tasks assigned to the network manager can be reassessed, allowing a focus on the more strategic and higher level aspects of network management.

The remainder of this paper is organized as follows. The next section discusses related work. The proposed system is then presented, detailing its main modules and features, as well as the technologies used in its development. A case study showing how the system can be deployed in practice is presented next. Finally, conclusions are drawn and future work is outlined.

\section{Related Work}\label{sec:related-work}

The detection of denial of service attacks in software-defined networks remains an active topic. Sawah \emph{et al.} \cite{sawah2025} evaluated five classical classifiers on the DDoS-SDN dataset, combining feature selection by backward elimination with hyperparameter tuning by \emph{Grid Search} and five-fold cross-validation, and reported an accuracy of 99.99\% for \emph{Random Forest}. The procedure is close to the one adopted in this work, although it is restricted to the classification stage, without executing countermeasures.

Elshewey \emph{et al.} \cite{elshewey2025} compared six deep learning models on SDN traffic previously balanced with SMOTE, with the best performance obtained by a hybrid CNN-GRU model. Rohith \emph{et al.} \cite{rohith2026} proposed a federated and explainable framework for SDN, with feature selection based on quantum-inspired particle swarm optimization. Both works prioritize classifier accuracy, whereas the present proposal emphasizes the automation of the complete cycle, from dataset creation to mitigation.

Regarding the reaction to the attack, Rajper \emph{et al.} \cite{rajper2026} presented a three-tier defense mechanism with port connection analysis for mitigation in SDN. Wu \emph{et al.} \cite{wu2025} proposed a cloud-native architecture that composes mitigation service chains on demand, with a genetic algorithm to decide their deployment. Yilmaz \emph{et al.} \cite{yilmaz2025} addressed network function sharing in slicing environments and proposed a slice-oriented DDoS filtering mechanism.

Lightweight and distributed approaches have also gained ground. Mendonça \emph{et al.} \cite{mendonca2025} showed that byte and packet counts obtained from off-the-shelf home routers, combined with a hierarchical Bayesian model that exploits space-time correlation, are sufficient to detect attacks close to their origin. Fan \emph{et al.} \cite{fan2025} explored contrastive learning in SDN-assisted federated learning environments, and Aboalela \emph{et al.} \cite{aboalela2025} combined feature pruning and deep learning for detection in IoT environments.

Recent literature therefore concentrates predominantly on the detection stage, with validation on public datasets. However, the integration into a single operational tool of the automated construction of the flow dataset, the assisted selection of the classifier, and the automatic triggering of blocking rules on the SDN controller remains little explored, and this is precisely the scope of the system described below.

\section{Proposed System}\label{sec:proposed-system}

The main focus in the development of the solution is to reduce the complexity perceived by the human administrator. From this perspective, the development of the system sought to absorb the complexity of the activities related to these processes, which are commonly performed manually \cite{safdar2018}. Automating attack diagnosis reduces the repair time, the probability of errors in the analysis and in the execution of countermeasures, and the operational cost \cite{ayoubi2018}.

The MVC (\emph{Model-View-Controller}) software architecture was used to develop the system, since it supports orthogonality and maintainability of the application source code. Python was chosen for the implementation because of its simple yet powerful syntax and its suitability for rapid prototyping, along with a wide range of libraries for data science and web development, among others.

\begin{figure}[H]
  \centering
  \includegraphics{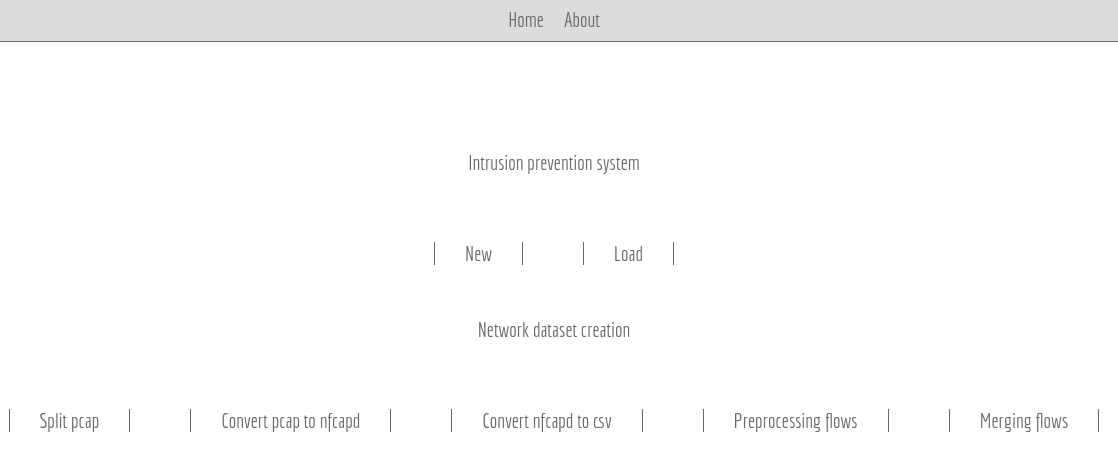}
  \caption{Home page of the system showing the two main modules: IPS and NDC.}
  \label{fig:home}
\end{figure}

The web application was developed with the Flask microframework and the Scikit-learn library, which supports the use of machine learning algorithms, data mining methods, and data analysis tools \cite{sklearn}. Other technologies, such as HTML, CSS, and JavaScript, were used to build the graphical user interface shown in Fig.~\ref{fig:home}. As can be seen in Fig.~\ref{fig:home}, the system was developed with two main modules, the \emph{Intrusion Prevention System} (IPS) and the \emph{Network Dataset Creation} (NDC), detailed in the next sections.

\subsection{Network Dataset Creation (NDC)}\label{subsec:ndc}

\begin{figure}[H]
  \centering
  \includegraphics[width=0.7\textwidth]{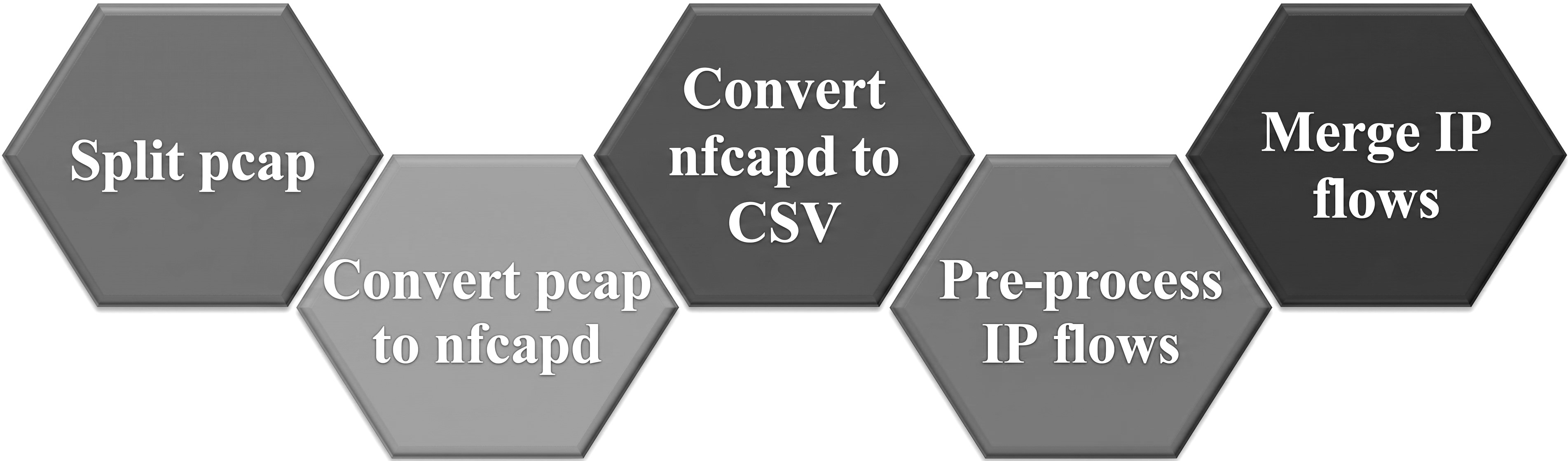}
  \caption{Main functions performed by the NDC module.}
  \label{fig:ndcfunc}
\end{figure}

The main goal of the NDC module is to obtain network data, analyze it, and create a dataset for training the machine learning algorithms. The data source used by NDC is the IP flow \cite{amaral2017}. One reason for using this data source is the reduction in the volume of data to be collected, processed, and stored, which shortens the analysis and decision time \cite{vaarandi2013}. NDC supports the creation of an IP flow dataset that accounts for the constant changes in the network traffic profile caused by new services, users, and applications that may appear in the monitored network. Fig.~\ref{fig:ndcfunc} presents the main functions performed by NDC.

The first function of NDC is responsible for splitting a \emph{pcap} file when it is too large to be processed. The \emph{tcpdump} tool is used in this stage. The second function converts such files to the \emph{nfcapd} format, which stores data as IP flows according to the NetFlow protocol. The \emph{nfpcapd} tool is used for this purpose. Since the \emph{nfcapd} format is binary, a third procedure is required to convert it to a CSV (\emph{Comma-separated values}) file, by means of the \emph{nfdump} tool. All tools used by the NDC module are open source.

The preprocessed IP flows stored in the CSV file are important to reach the final dataset used during model creation \cite{buczak2016}. During this research it was observed that many IP flow features did not contribute to machine learning, either because they were not representative or because they contained null values. Therefore, three main features of an IP flow were selected: duration in seconds (\emph{td}), number of packets (\emph{pkt}), and number of bytes (\emph{byt}), as shown in bold in Table~\ref{tab:flows}. Source and destination IP addresses were omitted because they are public addresses.

\begin{table}[H]
  \centering
  \caption{IP flows with the selected features in bold.}
  \label{tab:flows}
  \begin{tabular}{lllllllll}
  \toprule
  \textbf{sa} & \textbf{da} & \textbf{pr} & \textbf{flg} & \textbf{sp} & \textbf{dp} & \textbf{td} & \textbf{pkt} & \textbf{byt} \\
  \midrule
  ...165 & ...173 & udp & [0] & 58853 & 547 & \textbf{5} & \textbf{5} & \textbf{3648} \\
  ...165 & ...119 & tcp & [0, 1, 1, 0, 0, 1] & 54547 & 443 & \textbf{8} & \textbf{65} & \textbf{5420} \\
  ...165 & ...119 & tcp & [0, 1, 1, 0, 1, 1] & 54552 & 443 & \textbf{13} & \textbf{12} & \textbf{900} \\
  ...165 & ...119 & tcp & [0, 1, 1, 0, 1, 1] & 54551 & 443 & \textbf{13} & \textbf{11} & \textbf{868} \\
  \bottomrule
  \end{tabular}
\end{table}

\begin{figure}[H]
  \centering
  \includegraphics{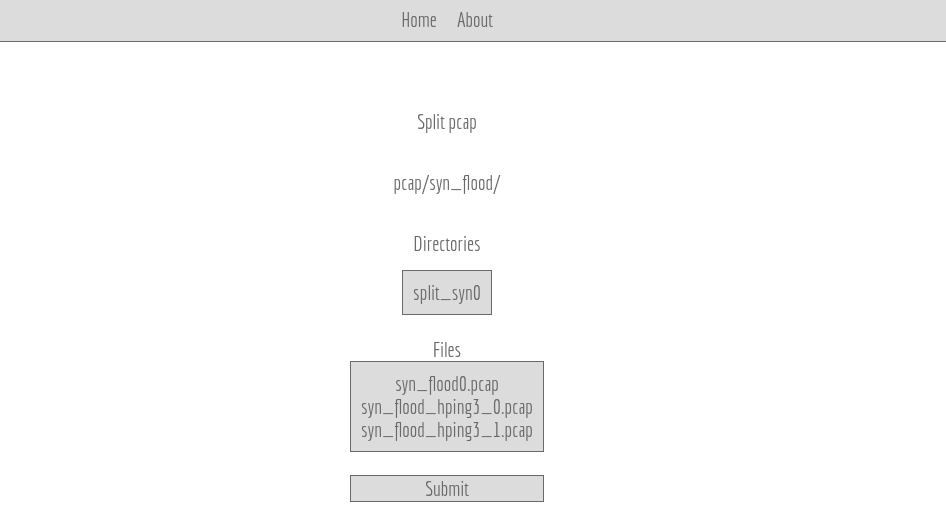}
  \caption{Graphical interface showing the operation of the NDC module.}
  \label{fig:ndcgui}
\end{figure}

During preprocessing, NDC aggregates the IP flows. This is done because previous analyses showed that aggregation reduces processing time and improves classifier results (\emph{e.g.}, accuracy). A method was therefore developed to aggregate flows based on the source IP address, the destination IP address, and the protocol, within a time interval defined by the user. A threshold parameter was established in the aggregation method in order to keep the aggregated dataset balanced in cases where thousands of flows would be reduced to a single sample. All these stages are performed automatically by the NDC module, as shown in Fig.~\ref{fig:ndcgui}.

\subsection{Intrusion Prevention System (IPS)}\label{subsec:ips}

\begin{figure}[H]
  \centering
  \includegraphics{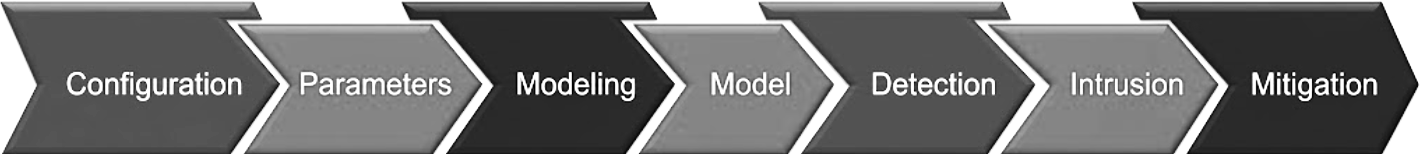}
  \caption{Overview of the main stages of the IPS module.}
  \label{fig:ipsstages}
\end{figure}

The main goal of the IPS module is to obtain the classification results of the machine learning algorithms and to trigger countermeasures in the software-defined network when an attack occurs. As shown in Fig.~\ref{fig:ipsstages}, this component performs four main operations: configuration, modeling, detection, and mitigation.

In the first stage, the network administrator can define the settings related to the machine learning algorithms. There are two options, creating a new model (\emph{New}) or loading a model saved on disk (\emph{Load}). If the second option is chosen, the system automatically starts monitoring and the procedures for attack detection and mitigation. Fig.~\ref{fig:ipsstages} summarizes these stages.

\begin{figure}[H]
  \centering
  \begin{minipage}[t]{0.48\textwidth}
    \centering
    \includegraphics[width=0.95\linewidth]{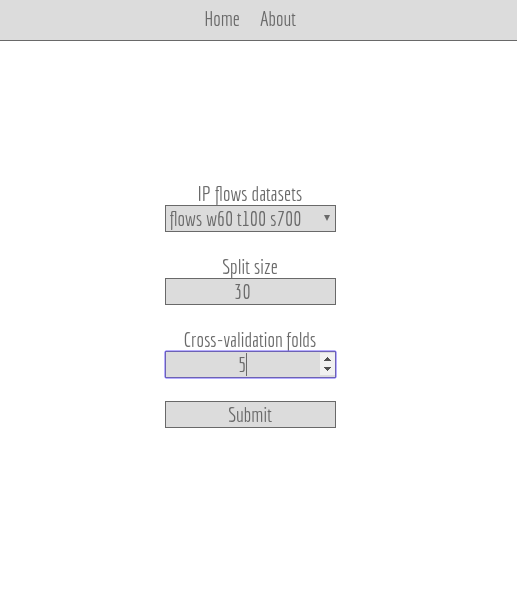}
    \caption{Initial configuration screen for creating a new model with the machine learning algorithms.}
    \label{fig:cfg1}
  \end{minipage}
  \hfill
  \begin{minipage}[t]{0.48\textwidth}
    \centering
    \includegraphics[width=0.95\linewidth]{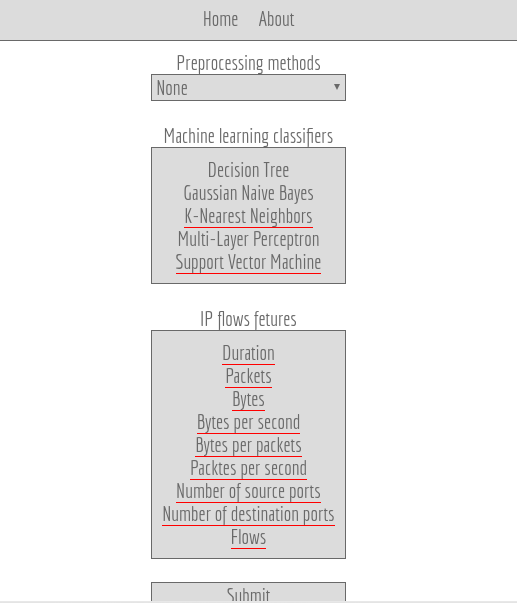}
    \caption{Selection of the machine learning algorithm and of the features used in training.}
    \label{fig:cfg2}
  \end{minipage}
\end{figure}

As shown in Fig.~\ref{fig:cfg1}, the system allows the user to choose which IP flow dataset will be used for training and testing the algorithms. It is possible to set the \emph{split size}, which is the size of the test set, and the \emph{cross-validation folds}, which is the number of subsets used for cross-validation. Other aspects of the configuration stage are presented in Fig.~\ref{fig:cfg2}, which allows the definition of the preprocessing methods (\emph{e.g.}, \emph{Min Max Scaler}), the machine learning algorithms, and the IP flow features used for learning.

In the current version, the system provides six machine learning algorithms: \emph{Decision Tree}, \emph{Random Forest}, \emph{Gaussian Naive Bayes}, \emph{K-Nearest Neighbors}, \emph{Support Vector Machine}, and \emph{Multi-layer Perceptron}. It should be noted that the system allows multiple algorithms to be selected, so that they can be modeled and later compared to determine which one is most suitable for the monitored network environment.

The second stage performed by the IPS module is modeling, which concerns the training of the algorithms and the optimization of hyperparameters \cite{buczak2016}. Some procedures commonly found in the literature were adopted for the chosen IP flow dataset. For example, if the value 30 is chosen as the split size, the data are randomly divided into a training set containing 70\% of the samples and a test set with the remaining 30\%.

This approach is important because evaluating the ability of a classifier on the training set would result in a biased \emph{score}; therefore, the test set is used to provide an unbiased estimate of the model ability \cite{brownlee2018}. Stratification is applied during this split, preserving in each part the same proportion of samples of each class as in the original dataset \cite{safdar2018}.

Machine learning algorithms have a set of hyperparameters, which are parameters not learned directly during training and which can affect the prediction and the computational performance of the models \cite{safdar2018}. The system was therefore developed to use \emph{Grid Search} together with \emph{Stratified n-fold cross-validation}, allowing it to identify the best combination of hyperparameters for each classifier, a strategy also adopted in recent SDN detection studies \cite{sawah2025}.

The \emph{cross-validation} technique is applied to the training set, splitting it according to the parameter \emph{k}, as shown in Fig.~\ref{fig:cfg1} (\emph{k} = 5). Similarly, this procedure sets aside an isolated subset to assess the ability of the model, preserving the proportion of each class during the splits. \emph{Grid Search} generates many combinations of the same model according to the defined hyperparameters.

\begin{figure}[H]
  \centering
  \includegraphics[width=0.85\textwidth]{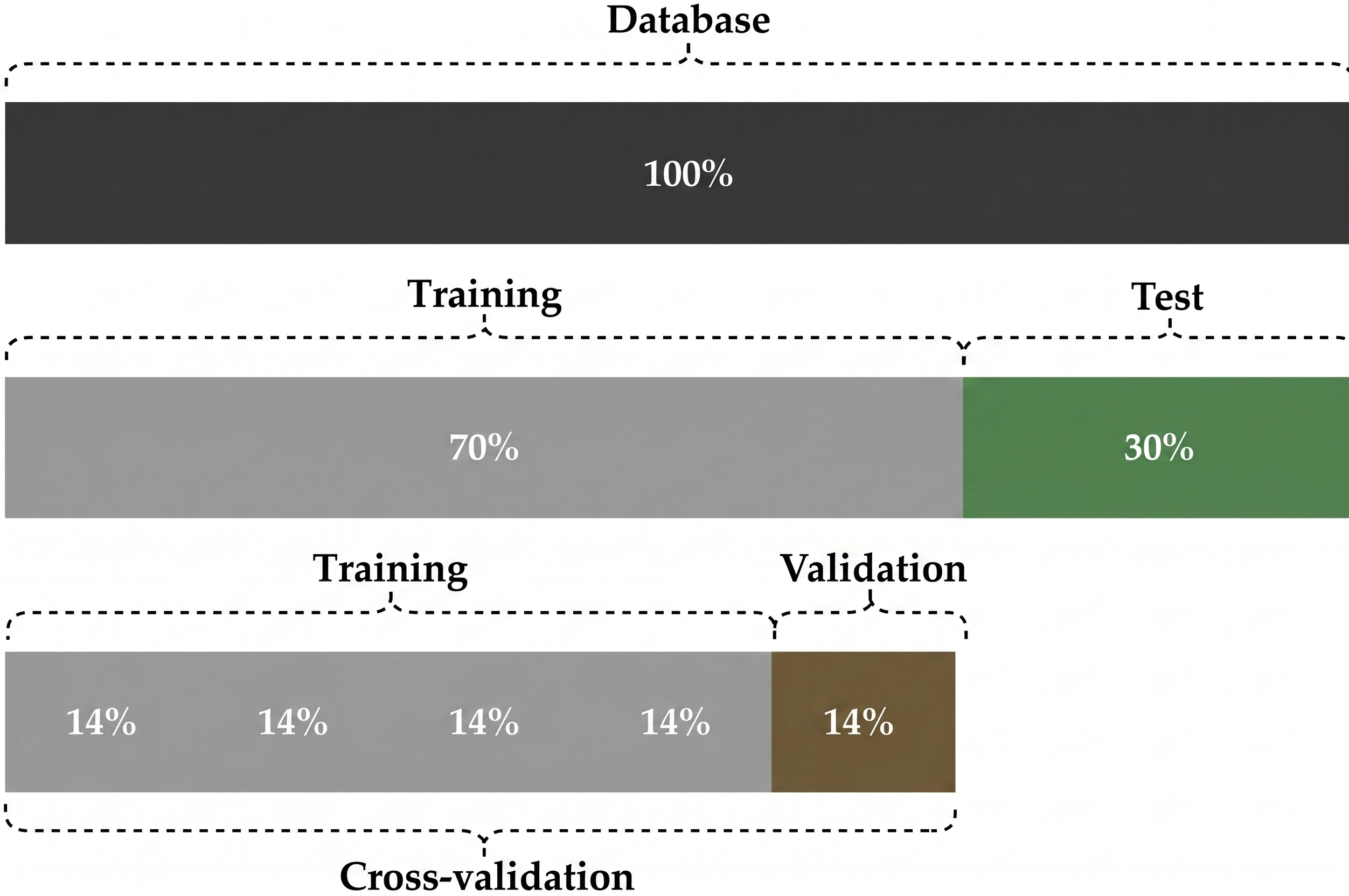}
  \caption{Splits applied according to the settings defined for the system.}
  \label{fig:splits}
\end{figure}

Each combination is trained on \emph{k-1} splits and evaluated on a validation set. This procedure is applied \emph{k} times in a row, alternating the order of the splits used. After the previous iteration ends, \emph{cross-validation} starts again for another combination, and the best result is chosen as the solution. Fig.~\ref{fig:splits} illustrates the splits applied to the datasets by the different techniques adopted.

\begin{figure}[H]
  \centering
  \includegraphics[width=0.65\textwidth]{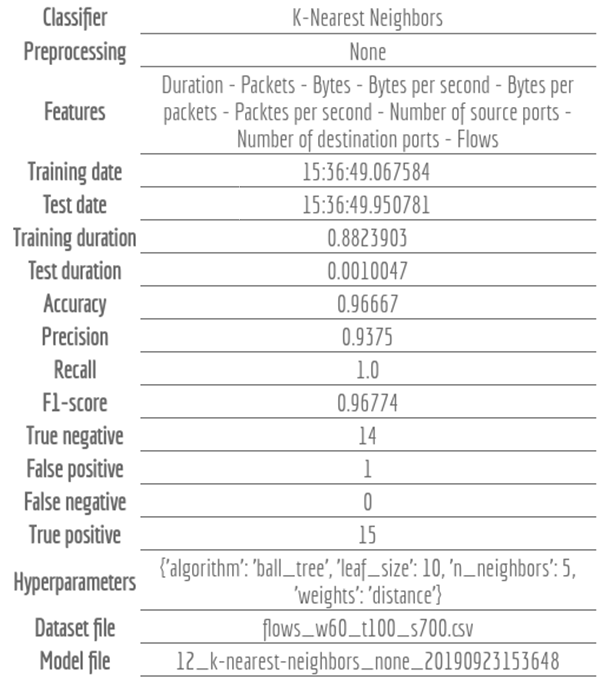}
  \caption{Information presented by the system regarding the training and testing of the machine learning algorithms.}
  \label{fig:results}
\end{figure}

After training the classifiers and selecting the hyperparameters, prediction is performed on the test set, which represents the data not observed during the training phase. The prediction results, which consist of the classes estimated by the algorithms, are used as the basis for the accuracy, precision, recall, and \emph{f1-score} metrics derived from the confusion matrix and defined in \cite{buczak2016}, \cite{hamid2016}. In addition, the developed system also presents other relevant information, as shown in Fig.~\ref{fig:results}.

In this stage, the system allows the network administrator to assess which algorithm achieved the best performance. After the selection, all data are stored and made available through the \emph{Load} option in Fig.~\ref{fig:home}, in order to start attack detection. The detection and mitigation stages of the IPS module are intended to detect attacks and apply repair actions automatically while an attack is in progress, and are discussed in the next section.

\section{Case Study and Results}\label{sec:case-study}

\begin{figure}[H]
  \centering
  \includegraphics[width=\textwidth]{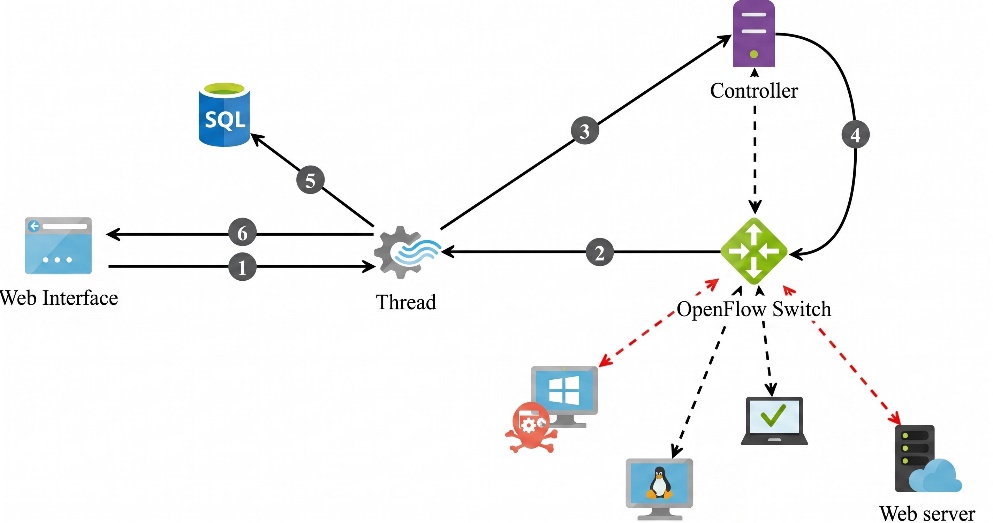}
  \caption{Overview of the use of the proposed system for detection and mitigation in an SDN.}
  \label{fig:environment}
\end{figure}

In order to validate the proposed system, a case study is presented here. The goal is to show the operation of the NDC and IPS modules in a software-defined network. An overview of the experimental environment is presented in Fig.~\ref{fig:environment}.

The NDC module is first invoked to carry out the training stage of the machine learning algorithms. For this purpose, a 95 GB network dataset from the university campus network was used \cite{lobato2016}, containing normal traffic and real network attacks divided into two main classes: \emph{Denial of Service} (DoS) and \emph{Probe}.

The attacks in the first class are ICMP flood, land, nestea, smurf, SYN flood, teardrop, and UDP flood. For network and system scanning, the following attacks were used: TCP SYN scan, TCP connect scan, SCTP INIT scan, Null scan, FIN scan, Xmas scan, TCP ACK scan, TCP Window scan, and TCP Maimon scan.

After generating the model and selecting the machine learning algorithm, the system, through the IPS module, starts stage 1 by launching a \emph{thread} responsible for collecting the network data (IP flows) exported in stage 2 by the OpenFlow switch at every interval \emph{n}. The experiments used \emph{n} = 60 s, which is a configurable parameter in the system. Upon receiving the IP flows, the NDC module performs preprocessing, extracts the features, and aggregates the flows.

If an attack occurs, the algorithm automatically blocks it upon detection through the IPS module, as shown in stages 3 and 4 in Fig.~\ref{fig:environment}. In these stages, the system pushes a blocking rule directly into the flow tables of the OpenFlow switch, through the REST API called \emph{Static Flow Entry Pusher} (SFEP) of the SDN controller. The Floodlight controller was used in the experiments. Once the attack is detected and resolved, various pieces of information (\emph{e.g.}, attack type, time, IP addresses involved) are stored in the database, as shown in stage 5. Stage 6 consists of presenting this information graphically to the network administrator.

The database was built with the sqlite3 library, which is an interface to the C library called SQLite. This technology provides a disk-based database, that is, a single file holds the entire structure and no separate process is required for its execution. In addition, a Python library called Flask SQLAlchemy was used, which supports the ORM paradigm and maps classes to database tables.

\begin{figure}[H]
  \centering
  \includegraphics{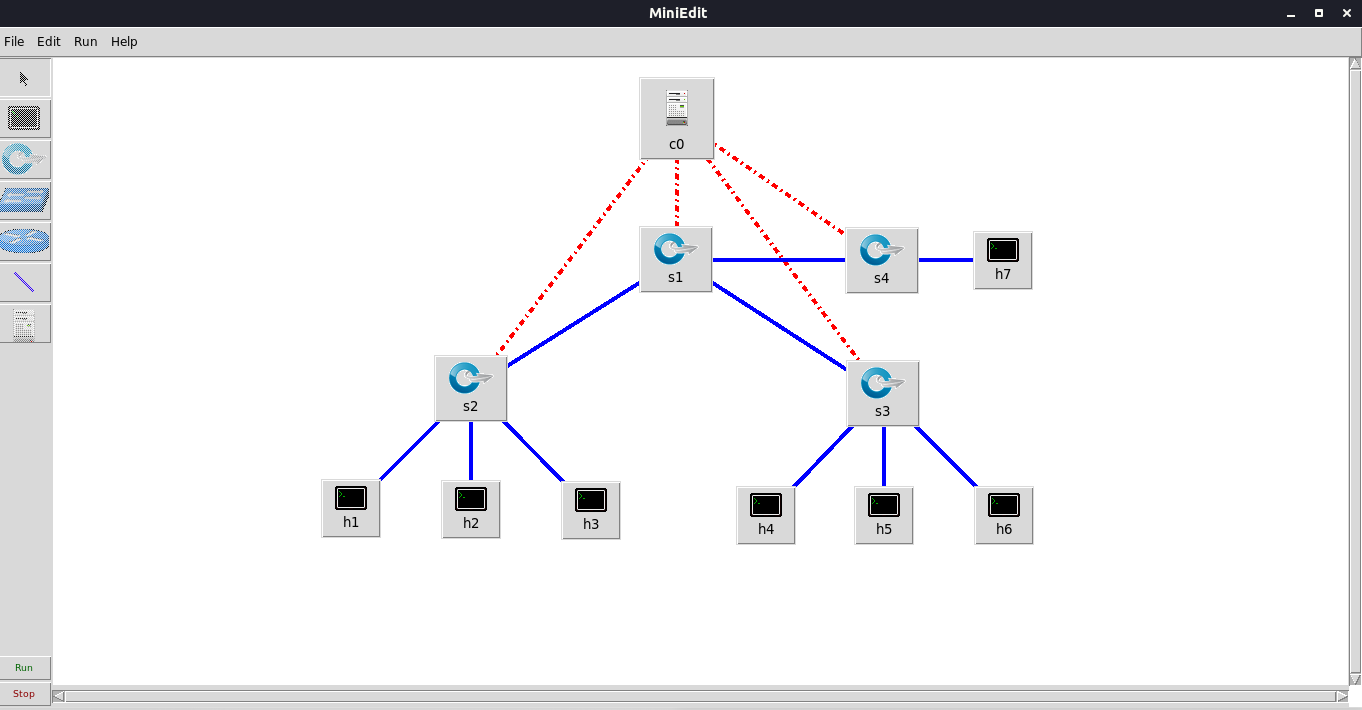}
  \caption{SDN network topology used in the experiments.}
  \label{fig:topology}
\end{figure}

A real example of a conducted experiment is presented here. A \emph{SYN flooding} DoS attack, using the Hping3 tool, was launched by \emph{host} H1 against a web server, H7. The SDN topology is presented in Fig.~\ref{fig:topology} and was created with Mininet through its Miniedit graphical interface.

\begin{figure}[H]
  \centering
  \begin{minipage}[t]{0.48\textwidth}
    \centering
    \includegraphics[width=0.95\linewidth]{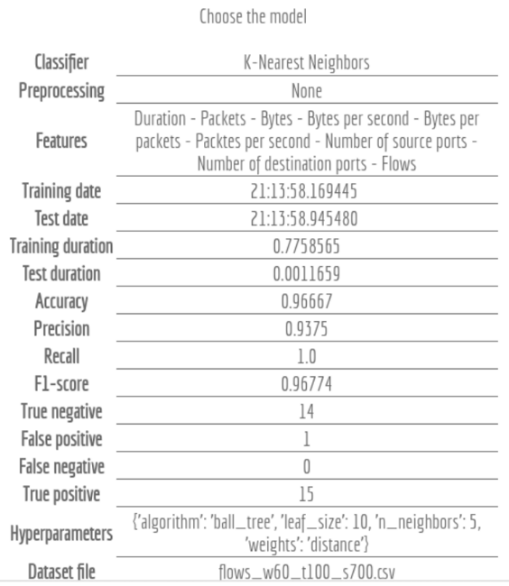}
    \caption{Selection of the KNN algorithm for detection.}
    \label{fig:knn}
  \end{minipage}
  \hfill
  \begin{minipage}[t]{0.48\textwidth}
    \centering
    \includegraphics[width=0.95\linewidth]{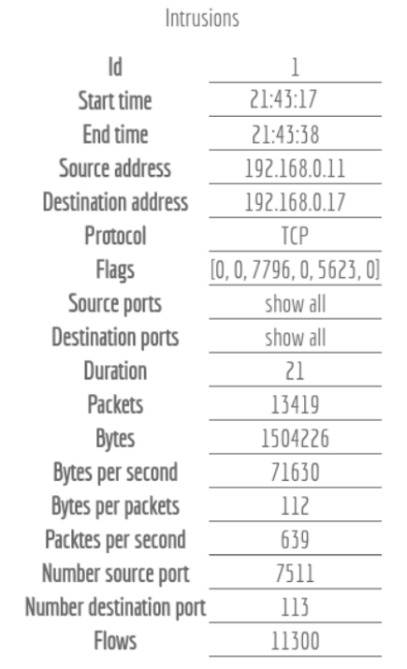}
    \caption{Information about the detection of the attack.}
    \label{fig:detection}
  \end{minipage}
\end{figure}

The KNN (\emph{K-Nearest Neighbors}) algorithm was used for detection. It was chosen because it showed good effectiveness for the network scenario used in the experiment. The main metrics used to select the algorithm were accuracy (96.6\%), precision (93.7\%), recall (100\%), and \emph{F1-score} (96.7\%), as shown in Fig.~\ref{fig:knn}.

\begin{figure}[H]
  \centering
  \includegraphics{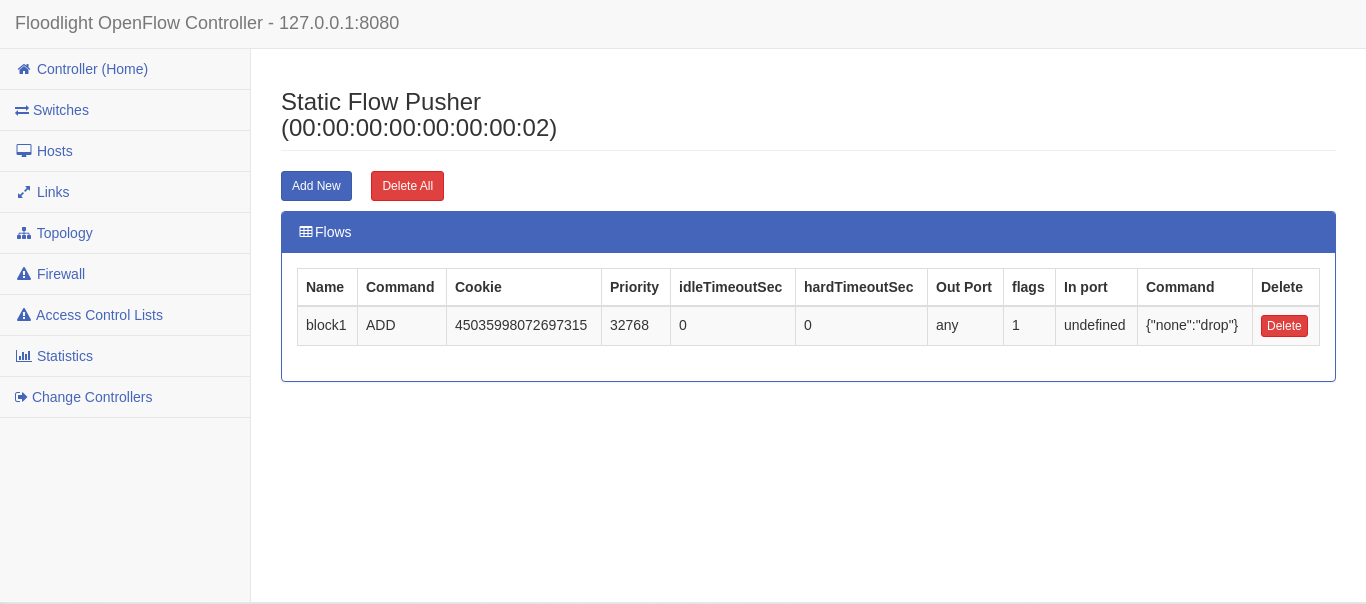}
  \caption{Blocking rule pushed by the system to the Floodlight controller to stop the attack from H1.}
  \label{fig:rule}
\end{figure}

As shown in Fig.~\ref{fig:detection}, the system successfully detected the attack and presents a summary of the event. The reported information includes the duration: the attack started at 21:43:17 and was interrupted at 21:43:38, that is, after 21 seconds the attack was automatically detected and blocked. Upon detecting the attack, the system automatically triggered a countermeasure to the Floodlight controller, which sent it to switch S2 in the attacker network (H1) so that blocking occurred instantly, thus stopping the attack. The blocking rule added to the SDN controller is presented in Fig.~\ref{fig:rule}.

\section{Conclusion}\label{sec:conclusion}

The challenges inherent to network management activities keep growing, in particular those aimed at ensuring data security and privacy. Automated systems that reduce manual operation are required, as well as the ability and flexibility to keep up with the evolution of networks and of attacks, which mutate constantly.

With breakout times now measured in minutes, an effective defense depends on two properties at once: a classifier accurate enough to be trusted without supervision, and a response path short enough to act within the attacker's window. The proposed system delivers both, since the same automated cycle that selects and evaluates the classifier also pushes the blocking rule to the controller, with no human operator in the critical path between detection and containment.

To address these challenges, this work presented an automated system that operates in real time and exploits the characteristics and advantages of machine learning algorithms and software-defined networks. The experiments and the case study presented here have shown its potential to automatically enable attack detection and mitigation with minimal human intervention.

As future work, other machine learning algorithms will be analyzed and implemented, and experiments will be carried out in large-scale networks. We also intend to compare the performance of the system with recent distributed and federated detection approaches, assessing the trade-off between accuracy and response time in production scenarios.

\begingroup
\sloppy

\endgroup

\end{document}